\documentclass[aip,apl,reprint,superscriptaddress,floatfix,amsmath,showpacs]{revtex4-2}
\usepackage{xcolor}
\usepackage{graphicx}
\usepackage{amsmath,bm,epsfig}
\definecolor{g-blue}{rgb}{0.83,0.95,1}
\usepackage{bm}
\usepackage{amssymb}
\usepackage{amsfonts}
\usepackage{amsmath}
\usepackage{graphicx}
\usepackage{ulem}
\usepackage{upgreek} 
\usepackage{textcomp} 
\usepackage[pdftex,colorlinks=true,allcolors=blue,breaklinks=true]{hyperref}  

\makeatletter
\def\set@curr@file#1{%
	\begingroup
	\escapechar\m@ne
	\xdef\@curr@file{\expandafter\string\csname #1\endcsname}%
	\endgroup
}
\def\quote@name#1{"\quote@@name#1\@gobble""}
\def\quote@@name#1"{#1\quote@@name}
\def\unquote@name#1{\quote@@name#1\@gobble"}
\makeatother

\def\Fbox#1{\vskip1ex\hbox to 8.5cm{\hfil\fboxsep0.3cm\fbox{%
			\parbox{8.0cm}{#1}}\hfil}\vskip1ex\noindent}  

\def\Sb#1{_{\scriptscriptstyle\rm{#1}}}
\def\He4 {$^4$He~}
\newcommand{\B}[1]{{\bm{#1}}}

\begin{document}
 	\title{Classical analog of qubit logic based on a magnon Bose–Einstein condensate}   
	 
	\author{Morteza~Mohseni}
	\email{mohseni@rhrk.uni-kl.de}
	\affiliation{Fachbereich Physik and Landesforschungszentrum OPTIMAS, Technische Universit\"{a}t Kaiserslautern, 67663 Kaiserslautern, Germany}
	
	\author{Vitaliy~I.~Vasyuchka}
	\email{vasyuchka@physik.uni-kl.de}
	\affiliation{Fachbereich Physik and Landesforschungszentrum OPTIMAS, Technische Universit\"{a}t Kaiserslautern, 67663 Kaiserslautern, Germany}

	\author{Victor~S.~L’vov}
	\email{victor.lvov@gmail.com}
	\affiliation{Fachbereich Physik and Landesforschungszentrum OPTIMAS, Technische Universit\"{a}t Kaiserslautern, 67663 Kaiserslautern, Germany}
	
	\author{Alexander~A.~Serga}
	\email{serga@physik.uni-kl.de}
	\affiliation{Fachbereich Physik and Landesforschungszentrum OPTIMAS, Technische Universit\"{a}t Kaiserslautern, 67663 Kaiserslautern, Germany}
	
	\author{Burkard~Hillebrands}
	\email{hilleb@physik.uni-kl.de}	
	\affiliation{Fachbereich Physik and Landesforschungszentrum OPTIMAS, Technische Universit\"{a}t Kaiserslautern, 67663 Kaiserslautern, Germany}
	
	\date{\today}
	
	\begin{abstract}
 
	We present a classical version of several quantum bit (qubit) functionalities using a two-component magnon Bose-Einstein condensate formed at opposite wavevectors in a room-temperature yttrium-iron-garnet ferrimagnetic film. The macroscopic wavefunctions of these two condensates serve as orthonormal basis states that form a system being a classical counterpart of a single qubit. Solving the Gross--Pitaevskii equation and employing micromagnetic numerical simulations, we first show how to initialize the system in one of the basis states: the application of wavevector-selective parallel parametric pumping allows us to form only a single condensate in one of the two lowest energy states of the magnon gas. Next, by translating the concept of Rabi-oscillations into the wavevector domain, we demonstrate how to manipulate the magnon-BEC system along the polar axis in the Bloch sphere representation. We also discuss the manipulation regarding the azimuthal angle.
	
	\end{abstract}
	
	
 	\maketitle

	There is an enormous need for faster and more efficient information processing. Quantum computing is widely discussed as a future computing technology, especially with regard to computing power and and the very favorable scaling properties \cite{arute2019IBM, PieterKok2007, Meyer2002, Nielsen2010, Williams2011, Ladd2010, Makhlin2001, Brien2008}. However, current quantum computing technologies are still far from ubiquitous. In particular, the need to operate in the milli-Kelvin temperature range is a significant obstacle.
	
\begin{figure*}
		\includegraphics[width=1.0\textwidth]{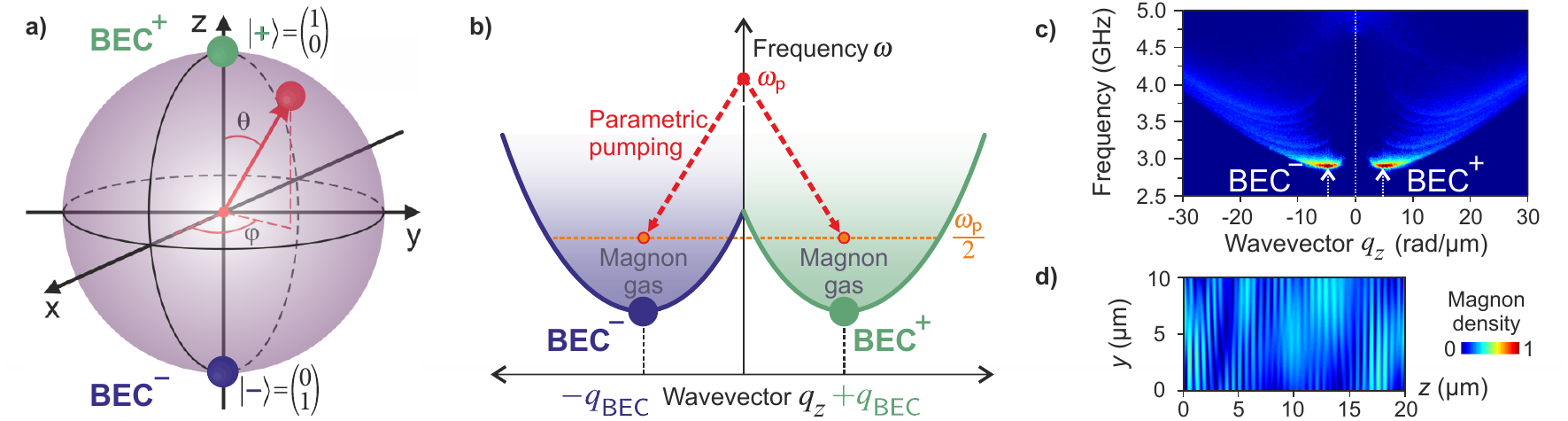}
		\caption{\textbf{System of two magnon BECs.} (\textbf{a})
		 The green and blue dots at the north and south pole of the Bloch sphere represent 
		pure  $|+\rangle$ and $|-\rangle$ states of $\mathrm{BEC}^{\mathrm{+}}$ and $\mathrm{BEC}^{\mathrm{-}}$ while the red dot shows the position of a mixed state of these magnon BECs.
		(\textbf{b}) Schematic spectrum of magnons and BEC generation via parametric pumping using $q \approx 0$ microwave photons. 
		(\textbf{c}) Numerical simulation of the condensation process of parametrically populated magnon gas in a YIG ferrimagnetic film. The fine structure of the magnon gas population corresponds to the thickness magnon modes of this film. The YIG film-thickness is 5\,$\upmu\mathrm{m}$. The pumping frequency is 7.5\,GHz with a pump duration of 50\,ns. The bias magnetic field is 100\,mT. d) Spatial interference pattern formed by the superposition of the $\mathrm{BEC}^{\mathrm{+}}$ and $\mathrm{BEC}^{\mathrm{-}}$ states.
		}
		\label{F:Concept}
	\end{figure*}

	In the field of quantum information technology, the basic unit of information is defined as a quantum bit or qubit for short \cite{Nielsen2010, Meyer2002, Deutsch2020, Chu-RyangWie2020, Hidary2019}. The qubit is most often represented by a superposition of two wavefunctions, which describe the two orthonormal basis states of the system. Thus, from the information content point of view, a cubit is a multi-valued object characterized by two independent continuous parameters, such as the ratio of wave function amplitudes and their relative phase. Qubits are regularly represented as states on the surface of a Bloch sphere, see Fig.\,\ref{F:Concept}(a). 
	
	In the search for physical systems suitable to represent a set of qubits, macroscopic quantum states of matter such as Bose--Einstein condensates (BEC) are promising candidates, in particular because of their inherent coherence and resistance to noise \cite{Makhlin2001, Qaiumzadeh2017, Tserkovnyak2017, Byrnes2012, Adrianov2014, Mohseni2021error, Semenenko2016, Xue2019polariton_qubit, Ghosh2020exciton-polariton_quantum-comp}. The BEC is a collective object, in which particles or quasiparticles represented by a coherent wavefunction populate a lowest energy state of the system \cite{Stone2015einstein, Einstein1924quantentheorie, Davis1995bose, Amo2009polaritons, Eisenstein2004excitons, Klaers2010photons, Bunkov2008BECinHe, Autti2021arXiv}. Such a BEC can be created using magnons in ferrimagnetic crystals, even at room temperature \cite{Demokritov2006bose, Nowik-Boltyk2012, Bugrij2013, Serga2014bose, Nakata2014persistent_current, Bozhko2016supercurrent, Bozhko2019Bogoliubov_waves, Safranski2017, Schneider2020rapid_cooling, Schneider2021rapid_cooling_SHE, Divinskiy2021, Kreil2021, Noack2021}.
	
    The work presented here addresses room-temperature, magnon-BEC-based qubit calculus. At room temperature, magnon BECs operate in the semi-classical regime, and some effects that exist only in the quantum mechanical regime, such as entanglement of states, are not accessible. However, a rather large subset of architectural aspects, algorithms, and \textit{modi operandi} used in quantum computing can be adapted and made operational. Many algorithms do not rely on quantum mechanical entanglement \cite{Biham2004, Lanyon2008} and can be implemented with the semi-classical qubit functionality discussed here. This approach removes the technical obstacles that exist in quantum calculations, such as the need to operate at milli-Kelvin temperatures, and allows potentially the operation in a room-temperature solid-state device with standard microwave interface technology.
    
    An important issue in this context is scaling. Quantum computing claims polynomial scaling of computation time with system size, which is very advantageous over the merely exponential scaling properties of conventional, classical Boolean logic. Recently, it has been shown that polynomial scaling holds for a number of algorithms that use wave-based computing in a non-quantum mechanical implementation (see, e.g., Refs.\,\cite{Balynsky2021, Cheng2021}). Thus, in terms of computation time, there is a fairly favorable gap between classical Boolean and true quantum-mechanical computing concepts, which can be filled by the magnon BEC concept presented here.

		\medskip
    {
	\fontsize{11pt}{10.0pt}
	\selectfont
	\noindent\textcolor{black}{\textbf{Results}}}

\medskip
\noindent \textbf{Magnon BEC as classical analog of a qubit.}
	The magnon BEC phenomenon addresses the spontaneous formation of a coherent state or wave---a macroscopic quantum state---in an otherwise disordered magnonic system. 
    At room temperature, a magnon BEC can form in a weakly interacting magnon gas if the chemical potential of the system is increased to the minimum magnon energy, for example by increasing the magnon density via external injection \cite{Demokritov2006bose, Serga2014bose, Safranski2017, Schneider2021rapid_cooling_SHE, Divinskiy2021}. The most prominent method to inject magnons is parallel parametric pumping \cite{Schloemann1962, NSW-book}, in which external microwave photons of frequency $\omega_\mathrm{p}$ and wavenumber $q_\mathrm{p} \simeq 0$ split into two magnons with the frequency $\omega_\mathrm{m}=\omega_\mathrm{p}/2$ and wavevectors $\pm \mathbf{q}_\mathrm{m}$ \cite{Melkov1999}. The overpopulated magnon gas near the bottom of the magnon spectrum relaxes to the two lowest energy states via multi-magnon scattering processes, forming a double-BEC state with the two BECs located in the two global minima of the system as shown in Fig.\,\ref{F:Concept}(b). The observed exciting dynamics of such a double-BEC, such as the inherent coherency \cite{Noack2021, Dzyapko2016}, interference patterns \cite{Nowik-Boltyk2012, Bugrij2013}, quantized vorticity\cite{Nowik-Boltyk2012}, supercurrents\cite{Nakata2014persistent_current, Bozhko2016supercurrent}, Bogoliubov waves\cite{Bozhko2019Bogoliubov_waves}, and Josephson oscillations\cite{Nakata2014persistent_current, Kreil2021}, all at room temperature, are further encouraging to consider magnon BECs for performing a classical subset of qubit computing operations.
	 
    We take into account the fact that for an in-plane magnetized film, in the two-dimensional in-plane wavevector landscape, the magnon frequency spectrum $\omega = \omega ( \B q )$ has two minima at symmetric values of the wavevectors $\pm \B q_{\mathrm{_{BEC}}}$, see Fig.\,\ref{F:Concept}(b). 
    Consequently, the condensate consists of two components, $\mathrm{BEC}^{\mathrm{+}}$ and $\mathrm{BEC}^{\mathrm{-}}$, forming the double-BEC and described by the wavefunctions	
    ${\mathrm{\Psi }}^+\left(x,t\right)\mathrm{exp}\left[i\left(q_{\mathrm{_{BEC}}}x-{\omega }_{\mathrm{_{BEC}}}t\right)\right]$ and 
    ${\mathrm{\Psi }}^-\left(x,t\right)\mathrm{\ exp}\left[i\left({-q}_{\mathrm{_{BEC}}}x-{\omega }_{\mathrm{_{BEC}}}t\right)\right]$,
    respectively. The fundamental basis for the description of the phenomena in this semi-classical limit is the nonlinear Schr\"{o}dinger equation (NSE), also known as Gross--Pitaevskii equation (GPE) \cite{Pitaevskii1998}. In the one-dimensional version, which is fully applicable here, it reads
\begin{subequations}\label{GP}  
 \begin{align}
  \label{GP_a} 
  \begin{split}
    \Big[i\frac{\partial }{\partial t} + D_{xx}\frac{{\partial }^2}{\partial x^2}
    & - T{\left|{\mathrm{\Psi}}^+\right|}^2 - S{\left|{\mathrm{\Psi }}^-\right|}^2 - P^+\left(x,t\right) \\ 
    & + i\mathrm{\Gamma }\Big] \cdot {\mathrm{\Psi }}^+\left(x,t\right)  =  iF^+\left(x,t\right)\, , 
  \end{split} \\
  \label{GP_b}
  \begin{split}
    \Big[i\frac{\partial }{\partial t}+D_{xx}\frac{{\partial }^2}{\partial x^2}
    & - T{\left|{\mathrm{\Psi }}^-\right|}^2-S{\left|{\mathrm{\Psi }}^+\right|}^2 - P^-\left(x,t\right) \\
    & + i\mathrm{\Gamma }\Big] \cdot {\mathrm{\Psi }}^-\left(x,t\right)  =  iF^-\left(x,t\right)\, . 
  \end{split}
 \end{align}
\end{subequations}
Here, $D_{xx}=\frac{1}{2}\frac{{\partial }^2\omega (\boldsymbol{q})}{\partial q^2_x}$ is the dispersion coefficient in the $x$-direction, and $T$ and $S$ are the amplitudes of self- and cross-interactions between $\mathrm{BEC}^{\mathrm{+}}$ and $\mathrm{BEC}^{\mathrm{-}}$, respectively. $P^{\pm }\left(x,t\right)$ are the external potentials caused, for instance, by variations of longitudinal magnetizing fields and $F^{\pm }\left(x,t\right)$ are the external forces caused, e.g., by transversal microwave magnetic fields. The magnon relaxation rate, originating, for example, from the interaction with the phonon bath, is denoted by $\mathrm{\Gamma }$. For $\mathrm{\Gamma}=0$, Eq.\,\eqref{GP} has stationary solutions, which describe the two magnon BECs at the bottom of the spin-wave spectrum in the $+q_\mathrm{_{BEC}}$ and $-q_\mathrm{_{BEC}}$ positions. If $\mathrm{\Gamma }\neq 0$, the wavefunctions ${\mathrm{\Psi }}^{\pm }\left(x,t\right)$ decay as $\mathrm{exp}\left(-\mathrm{\Gamma }t\right)$.

In the space-homogeneous case, there is, apart from the oscillating factor $\mathrm{exp}[i(q_{\mathrm{_{BEC}}}x)]$ no $x$-dependence, and Eq.\,\eqref{GP} becomes identical to the general form of a degenerate two-level system. An example of such a system is a quantum particle with spin $S=\frac 12$, which has a wavefunction consisting of a spin-up ${\psi }^+\Sb{1/2}$ and a spin-down ${\psi }^-\Sb{1/2}$ component. To stress this analogy, we introduce the normalized BEC wavefunctions
\begin{subequations}\label{Norm}  
 \begin{align}
  \label{Norm_a} 
  \begin{split}
{\psi }^{\pm }\left(x,t\right)={{\mathrm{\Psi }}^{\pm }\left(x,t\right)}/{\sqrt{{\left|{\mathrm{\Psi }}^+\left(x,t\right)\right|}^2+{\left|{\mathrm{\Psi }}^-\left(x,t\right)\right|}^2}} \, ,
  \end{split} \\
  \label{Norm_b}
  \begin{split}
{\left|{\psi }^+\left(x,t\right)\right|}^2+{\left|{\psi}^-\left(x,t\right)\right|}^2=1\, .
  \end{split}
 \end{align}
\end{subequations}

The qubit state position on the Bloch sphere [see Fig.\,\ref{F:Concept}(a)] is given by the polar angle $\theta $ and azimuthal angle $\varphi $. In our case, they are defined by the normalized densities of the two BECs, $N^{\pm }={\left|{\psi }^{\pm }\left(x,t\right)\right|}^2$ with
\begin{subequations}\label{NormDen}  
 \begin{align}
  \label{NormDen_a} 
  \begin{split}
     N^+ \left( x,t \right) + N^- \left( x,t \right) = 1 \, , 
  \end{split} \\
  \label{NormDen_b} 
  \begin{split}
      \theta = \mathrm{Arccos} \left\{ \left[ N^+ \left( x,t \right) - N^- \left( x,t\right) \right]/2 \right\} \, ,
  \end{split} \\
  \label{NormDen_c}
  \begin{split}
       \varphi =\mathrm{Arg}\left\{ {\psi }^+\left(x,t\right)\cdot {\psi }^{-*}\left(x,t\right) \right\} \, .
  \end{split}
 \end{align}
\end{subequations}
Thus, to address any position on the Bloch sphere, we need to control i) the number of magnons in each of the two BECs and ii) their relative phase. The implementation of this control will be demonstrated below by means of numerical micromagnetic simulations \cite{Mumax, Mohseni2020}. 

\begin{figure*}  
		\includegraphics[width=1.0\linewidth]{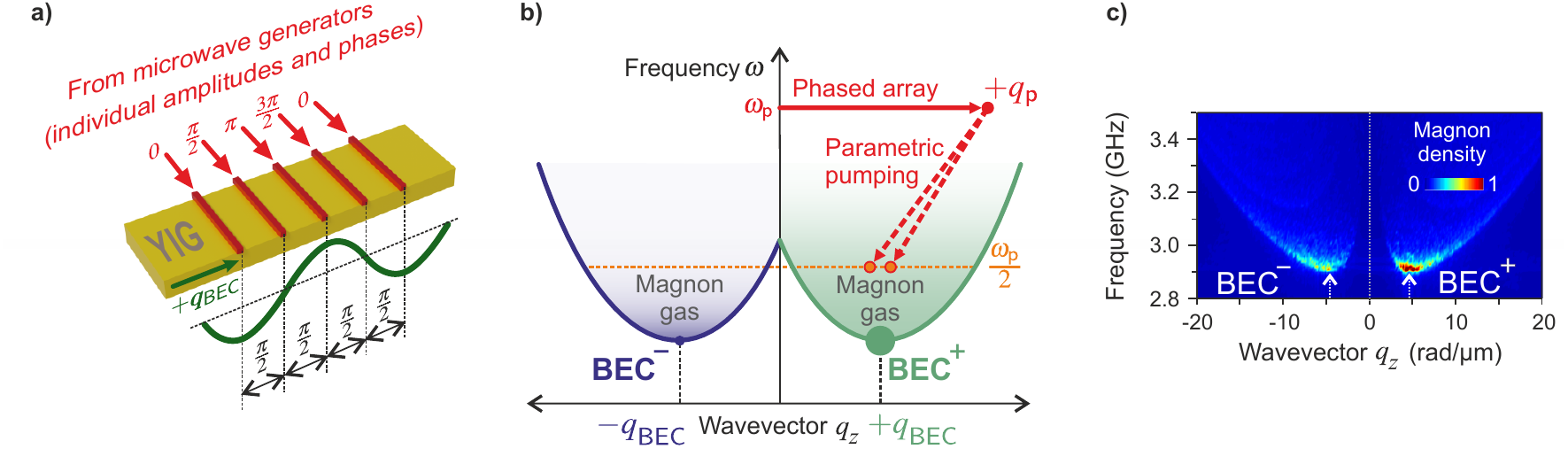}
		\caption{\textbf{Wavevector-selective parametric pumping and formation of an asymmetric magnon BEC.} 
		(\textbf{a}) Phased-array magnetic field excitation mechanism. 
		(\textbf{b}) Schematic view of wavevector-selective parametric pumping scheme. 
		(\textbf{c}) Micromagnetic modeling of BEC formation in a parametrically populated magnon gas. The color code indicates the normalized magnon density.  
			}
		\label{F:Asymmetric_pumping}
\end{figure*}

\medskip
\noindent \textbf{Initialization of magnon double-BEC state at the equator of the Bloch sphere.} A convenient way to generate a magnon BEC is to use the parametric excitation mechanism by a space-homogeneous electromagnetic field for generation of magnons in the gaseous magnon regime and to let these magnons condense into the two BEC states via dominating four-magnon-scattering events \cite{Demokritov2006bose,Serga2014bose}. Due to the momentum conservation during the magnon termalization processes, the resulting numbers of magnons in the $\mathrm{BEC}^{\mathrm{+}}$ and $\mathrm{BEC}^{\mathrm{-}}$ states also remain the same, $N^{\mathrm{+}} = N^{\mathrm{-}}$, as it has been evidenced by numerous magnon BEC experiments \cite{Nowik-Boltyk2012, Bozhko2016supercurrent, Bozhko2019Bogoliubov_waves, Kreil2021} and by results of numerical modeling shown in Fig.\,\ref{F:Concept}(c). It means that such a prepared initial double-BEC state lies somewhere on the equator of the Bloch sphere, $\cos\theta = 0$, and the azimuthal angle $\phi$ has an arbitrary value (further below we will address how to control $\phi\mathrm{}$). We note that the coherency of the generated BECs is evidenced by the spatial interference pattern of the two condensates as observed experimentally \cite{Nowik-Boltyk2012}, and shown by our micromagnetic simulations displayed in Fig.\,\ref{F:Concept}(d). Despite the presence of small non-uniformities in the spatial interference pattern evidencing the existence of quantized vorticities, the coherency of the two BECs on the larger scale is preserved \cite{Nowik-Boltyk2012, Rezende2009, Noack2021}. 

\medskip
\noindent \textbf{Initialization of magnon double-BEC state at the poles and any latitude of the Bloch sphere.} We propose to use a moving periodic pattern of a microwave magnetic field in order to pump magnon pairs dominantly to the positive or negative half of the wavevector space. It acts like an artificial propagating microwave field with controlled wavevector and frequency. This moving pattern can be created by an array of \textit{L} parallel wires separated by a distance $a$ and connected to a set of phase-locked microwave generators operating at the pumping frequency $\omega_\mathrm{p}$, see Fig.\,\ref{F:Asymmetric_pumping}(a). They work analogous to a phased array \cite{Visser2006, Visser2005}: The phases of the signals applied to the wires are chosen such that the phase difference $\mathrm{\Delta}_\mathrm{wire}$ between neighbored wires is fixed to $\mathrm{\Delta }_\mathrm{wire}=2\pi / K$ with $K$ being an integer number. Thus, moving magnetic field patterns are generated, which oscillate with $\omega_\mathrm{p}$ and propagate with wavevector $|\B {q}_\mathrm{p}|=2\pi/(Ka)$. For $K>2$ the field pattern has a well-defined unique direction of propagation, which is indicated by the sign of $\mathrm{\Delta }_\mathrm{wire}$. 

Such a field pattern, which is characterized by a reciprocal lattice vector $\left|\B {q}_\mathrm{p}\right|$ pointing into the direction of motion and the microwave frequency $\omega_\mathrm{p}$, will pump pairs of co-propagating magnons with frequencies $\omega_\mathrm{p} /2$ and wavevectors $q_\mathrm{pm}\le q_\mathrm{p}/2,\ \B {q}_\mathrm{pm} \parallel \B {q}_\mathrm{p}$,  see Fig.\,\ref{F:Asymmetric_pumping}(b). In this scenario, the thermalization of the parametric magnons will result in the population of either $\mathrm{BEC}^\mathrm{-}$ or $\mathrm{BEC}^\mathrm{+}$ states. This means that the prepared initial state of the magnon double-BEC is either at the north or south pole of the Bloch sphere. By proper choice of $\mathbf{q}_\mathrm{p}$, it is possible to populate $\mathrm{BEC}^\mathrm{+}$ and $\mathrm{BEC}^\mathrm{-}$ states with predefined densities $N^\mathrm{+}$ and  $N^\mathrm{-}$, i.e., to adjust the polar angle of the magnon system. The validity of this approach is demonstrated by the results of our micromagnetic simulations as shown in Fig.\,\ref{F:Asymmetric_pumping}(c).
 
 \begin{figure*}  
	\includegraphics[width=1.0\linewidth]{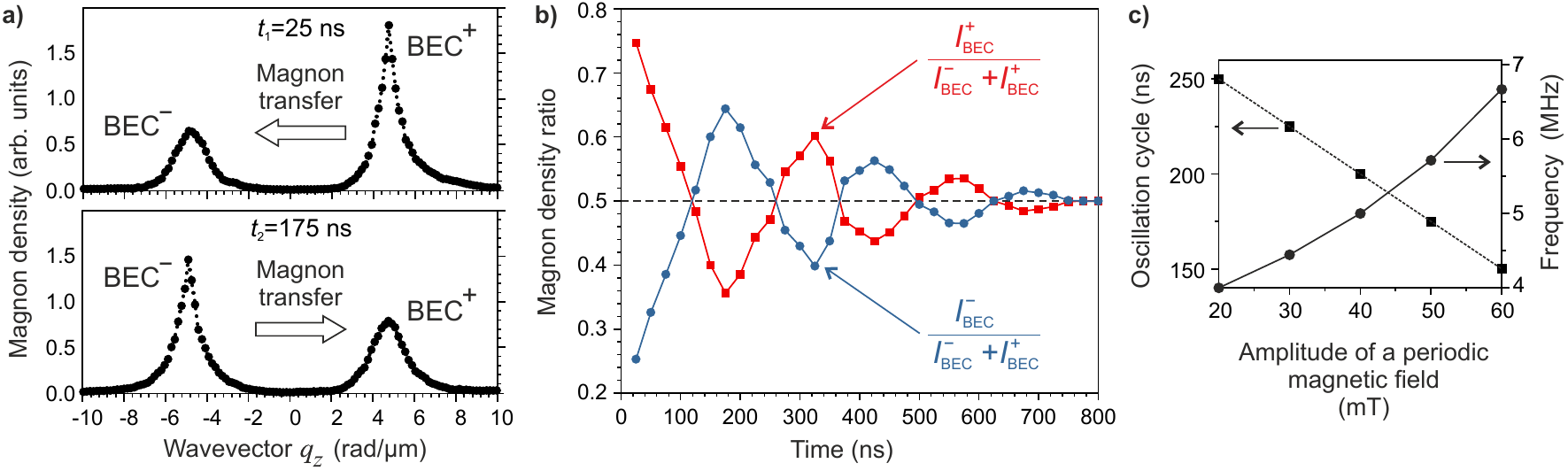}
	\caption{\textbf{Rabi oscillations in magnon BEC within a magnonic crystal.}
	(\textbf{a}) Wavenumber-dependent distribution of the magnon density at the bottom of spin-wave spectrum for two moments of time. 
	(\textbf{b}) Normalized peak magnon densities at $+q_{_\mathrm{BEC}}$ and $-q_{_\mathrm{BEC}}$ as functions of time.
	(\textbf{c}) Periodicity of the Rabi oscillations as a function of the magnitude of a spatially-periodic magnetic field induced by the dynamic magnonic crystal. 
	}
	\label{F:Rabi}
\end{figure*}

\medskip
\noindent \textbf{Semi-classical qubit protocols using Rabi-like oscillations}. Achieving a controlled change in the state of the previously initialized double-BEC system, e.g., a transformation of $\mathrm{BEC}^{\mathrm{+}}$ into $\mathrm{BEC}^{\mathrm{-}}$ and vice versa, and addressing any intermediate state, is essential for computing.  Indeed, our objective is to realize unitary transformations on the Bloch sphere to resemble logic state operations. 

The following tools and approaches serve as key ingredients for implementing the functionalities of the Pauli-X and -Z gates and the Hadamard gate. For this purpose, we use the concept of Rabi-oscillations, which refers to a cyclic energy exchange in a two-level quantum system in the presence of a driving field \cite{Griffiths, Dudin2012, Rosanov2013}. The system of the two magnon BECs represents the analogy of a two-level system in the wavevector domain, and thus, we translate the Rabi cycle into this domain as discussed in the following.

Let us consider the double-BEC system initialized in the north pole state, i.e., at $\mathrm{BEC}^{\mathrm{+}}$ as shown in Fig.\,\ref{F:Asymmetric_pumping}(c).
Instead of a time-dependent field, here we use a space-dependent stationary magnetic field to induce a Rabi cycle between the two BEC wavevector components. This field is generated using a dynamic magnonic crystal (MC) \cite{Chumak2010} similar to the setup that is shown in Fig.\,\ref{F:Asymmetric_pumping}(a). However, instead of a phased array driven by microwave currents, we insert pulsed DC currents into the wires to control the bias magnetic field. Consequently, this changes the energy landscape of the system. Once this MC is turned on, it creates a gap in the magnon dispersion relation at the wavevector corresponding to the periodicity of the crystal. This leads to a Bragg reflection of the incoming magnons at that particular wavevector, which in our case is $q_{_\mathrm{BEC}}$ \cite{Chumak2010}. However, if the magnons already exist in the system and the MC is switched on, geometrically-induced coupling between the existing magnons and the MC leads to the oscillations of the magnons between the opposite wavevector-states given their equivalent frequencies \cite{Karenowska2012}.

In a numerical simulation experiment, we activate the MC once the pure $\mathrm{BEC}^\mathrm{+}$ state is established. Figure\,\ref{F:Rabi}(a) shows the distribution of the condensed magnon density at the wavevector space in two given time spans. During the course of time from, e.g., $t_1 = 25$\,ns to $t_2 = 175$\,ns, the magnons are transferred from $\mathrm{BEC}^\mathrm{+}$ to $\mathrm{BEC}^\mathrm{-}$. The oscillation cycle is more visible in Fig.\,\ref{F:Rabi}(b), which represents the normalized magnon densities at $\mathrm{BEC}^\mathrm{-}$ and $\mathrm{BEC}^\mathrm{+}$ as a function of time. Interestingly, the oscillations between the two condensates continue until the condensed magnons dissipate due to the magnetic damping. Moreover, the oscillation cycle can be tuned using the MC modulation amplitude, i.e., by the amplitude of the spatially periodic magnetic field produced by the MC. As shown in Fig.\,\ref{F:Rabi}(c),
increasing the MC amplitude leads to shorter oscillation cycles, and consequently higher oscillations frequencies. This can be understood by a stronger coupling of the magnons and the MC \cite{Karenowska2012}. 
Indeed, the repetition of this action turns the double-BEC back into its initial state, thus representing the bit-flip operation described by the Pauli-$X$ matrix, $X= \big[\begin{smallmatrix}
0 & 1\\
1 & 0
\end{smallmatrix}\big]$. 
By changing the oscillation cycles, it is also possible to transfer the system to a superposition of $\mathrm{BEC}^\mathrm{+}$ and $\mathrm{BEC}^\mathrm{-}$, characterized by finite densities of both BECs. The proposed and demonstrated mechanism not only further suggests the coherency of the generated condensates, but also provides a strong tool to manipulate the magnon BEC states on the Bloch sphere even as a function of time, which is an additional degree of freedom for this purpose.

Furthermore, controlling the azimuthal angle of the double-BEC system is essential as well. This is possible by controlling the relative phase between the two BECs. By inserting a low-frequency microwave RF field with a linewidth smaller than the linewidth of the BEC into a structure similar to the one used to realize the dynamic MC, it is possible to change the relative BEC phase and consequently the position of the double-BEC system on the Bloch sphere. We emphasize that patterning such structures is a task that can be accomplished with today's nanostructuring techniques. 

The dynamic MC can be optimized for simultaneous work with phase-adjusted RF and pulsed DC-signals. This allows one to manipulate both the polar and azimuthal angles of a double-BEC state, and, by extension, to pave the way towards implementation of various phase-shift gates. For example, as an outlook, we expect the implementation of the Pauli-$Z$ gate, represented by the Pauli-$Z$ matrix, $Z=
\big[\begin{smallmatrix}
1 & 0\\
0 & -1
\end{smallmatrix}\big]$. 
By having the possibility of controlled rotation in relation to the different axes of the Bloch’s sphere, the functionality of the Hadamard gate,  $H=\frac{1}{\sqrt{2}}
\big[\begin{smallmatrix}
1 & 1\\
1 & -1
\end{smallmatrix}\big]$ 
can be achieved. 

	\medskip
    {
	\fontsize{11pt}{10.0pt}
	\selectfont
	\noindent\textcolor{black}{\textbf{Discussion}}}

\noindent The aim of the presented study is to demonstrate that several qubit calculus functionalities are feasible to implement using a magnon double-BEC at room temperature. Our work might initiate a new direction in the field of magnonics by bridging the field of macroscopic quantum states of magnons and quantum computing functionalities \cite{Elyasi2020, Tabuchi2015, Quirion2020}. We have used two spatially overlapping and interfering stationary magnon BECs with opposite wavevectors as a classical analog of a qubit. We presented the wavevector-selective parallel pumping mechanism for initialization of this double-BEC system  at the north or south pole of the Bloch sphere. Moreover, we have shown that Rabi-like oscillations can be translated into the wavevector domain to manipulate the magnon double-BEC state on the Bloch sphere. Realistic numerical simulations close to the experimental conditions provide evidence for the feasibility of this approach and they also shed light on the proposed mechanisms.

As an outlook, the next steps to be taken are obvious: It is the realization of coupling of two and more double-BEC magnon systems. The coupling of spatially separate magnon BECs has already been demonstrated, leading to the phenomenon of magnon Josephson oscillations \cite{Kreil2021, Autti2021, Fripp2021}. Microwave antenna structures for the phase coherent detection and generation of magnon BECs can be used to coherently couple double-BEC systems over larger spatial distances.

In summary, our study provides a significant momentum in the frontier between the functionality of a quantum bit (qubit) that make substantial use of entanglement, and systems consisting of an interacting set of classic BECs implementing a subset of qubit-based algorithms for which entanglement is not required. 

	\medskip
    {
	\fontsize{11pt}{10.0pt}
	\selectfont
	\noindent\textcolor{black}{\textbf{Methods}}}
	
{   \fontsize{9pt}{12.0pt}
	\selectfont

\noindent{\textbf{Micromagnetic modeling.}} Solving the equation of the magnetization motion have been carried out numerically using the open source MuMax 3.0 package. It uses the Dormand-Prince method \cite{Mumax} for the integration of the Landau-Lifshitz-Gilbert (LLG) equation:
\begin{equation}\label{LLG}  
      \frac{d\mathbf{M}}{dt} = -|\gamma| \mathbf{M} \times \mathbf{B}_\mathrm{eff} + \frac{\alpha}{M_\mathrm{s}} \left( \mathbf{M} \times \frac{d\mathbf{M}}{dt} \right) \, ,
\end{equation}
where $\mathbf{M}$ is the magnetization vector, $M_\mathrm{s}$ is the saturation magnetization, $\mathbf{B}_\mathrm{eff}$ is the effective magnetic field, $\gamma$ is the gyromagnetic ratio, and $\alpha$ is the damping constant. The system under investigation is a film with dimensions of 51.2\,\textmu m × 25.6\,\textmu m × 5\,\textmu m that is divided into 1024 × 512 × 16 cells. An external magnetic film is applied in the plane with an amplitude of 100\,mT. Realistic magnetic parameters of the YIG film are used: $M_\mathrm{s}$ = 140\,kA/m, $\alpha$ = 0.0002, the exchange constant $A_\mathrm{exch}$ = 3.5\,pJ/m \cite{Cherepanov1993saga_YIG, Mohseni2020, Rezende2009}. Magnons are injected using parallel parametric pumping, in which, microwave photons of frequency of $\omega_\mathrm{p}$  = 7.5\,GHz generate magnons with the frequency $\omega_\mathrm{p}$/2 = 3.75\,GHz. The duration of the pumping pulse is fixed to $t_\mathrm{pump}=50$\,ns in all simulations, and the system is then relaxed toward its equilibrium state to permit the BEC to establish at the bottom of the magnon spectrum. All simulations have been carried assuming room temperature. The used setup in this study resembles typical conditions of real experiments \cite{Demokritov2006bose, Serga2014bose}. 

The micromagnetic simulations were carried out in two steps. First, the external field is applied in the film plane to establish the relaxed magnetization state. This state is consequently used as the ground state in the dynamic simulations, in which parallel parametric pumped magnons are generated using an oscillating field that is applied parallel to the static field. The dynamic components of the magnetization, $M (x,y,z,t)$ of each cell are collected over a period of time. To process the raw data, one and two-dimensional fast Fourier transformation (FFT) has been carried out on the collected data set \cite{Mohseni2020}.
}

	\medskip
    {
	\fontsize{11pt}{10.0pt}
	\selectfont
	\noindent\textcolor{black}{\textbf{Data availability}}}
	
\noindent The data that support the findings of this study are available from the corresponding author upon request.

	\medskip
    {
	\fontsize{11pt}{10.0pt}
	\selectfont
	\noindent\textcolor{black}{

	\medskip
    {
	\fontsize{11pt}{10.0pt}
	\selectfont
	\noindent\textcolor{black}{\textbf{Acknowledgments}}}
	
\noindent This research was funded by the European Research Council within the Advanced Grant No.\,694709 SuperMagnonics and by the Deutsche Forschungs\-gemein\-schaft (DFG, German Research Foundation) within the Transregional Collaborative Research Center---TRR 173/2---268565370 ``Spin+X'' (projects B01, B04). The authors are grateful to Dmytro A. Bozhko for fruitful discussions.

	\medskip
    {
	\fontsize{11pt}{10.0pt}
	\selectfont
	\noindent\textcolor{black}{\textbf{Author contribution}}}

\noindent M.M. performed the numerical simulations, analyzed the data. V.V., V.L. and A.S. devised and planned the project and analyzed the general analytical model. B.H. led the project. All authors discussed the results and contributed to writing the manuscript.


\end{document}